\documentclass[preprint]{revtex4}

\usepackage{graphicx}
\usepackage{dcolumn}
\usepackage{bm}
\usepackage{subfig}

\def\be{\begin{equation}}
\def\ee{\end{equation}}
\def\bea{\begin{eqnarray}}
\def\eea{\end{eqnarray}}
\def\dprime{\prime\prime}
\def\calh{\mathcal{H}}

\begin{document}

\title{Quantum Remnant as Dark Energy and Dark Matter}
\author{Sang Pyo Kim}\email{sangkim@kunsan.ac.kr}
\affiliation{Department of Physics, Kunsan National University,
Kunsan 573-701, Korea} \affiliation{Asia
Pacific Center for Theoretical Physics, Pohang 790-784, Korea}

\author{Seoktae Koh}\email{skoh@itp.ac.cn}
\affiliation{Institute of Theoretical Physics, Chinese Academy of
Sciences, P.O. Box 2735, Beijing 100190, China}


\begin{abstract}
We study the quantum remnant of a scalar field
 protected by the uncertainty
principle. The quantum remnant that survived the later
stage of evolution of the universe may provide dark energy and dark
matter depending on the potential. Though the quantum remnant shares
some useful property of complex scalar field (spintessence)
 dark energy  model,
quantum fluctuations are still unstable to the linear perturbations
for $V \sim \phi^q$ with $q<1$ as in the spintessence model.
\end{abstract}

\maketitle

The current observational data predicts that the present universe
has recently been dominated by dark energy and dark matter in the
latest evolution \cite{Perlmutter:1998np, Komatsu:2008hk}. The
origin and property of dark energy and dark matter
 is one of the most challenging problems of this century. There have
been many models and proposals to explain the nature of dark energy
 \cite{Peebles:2002gy} and dark matter \cite{Bertone:2004pz}.

One of such attempts is to introduce a complex scalar field,
equivalent to two real scalar fields with the same coupling
constants, whose conserved charge leads to a centrifugal force to
the potential that may provide dark energy, dubbed spintessence
\cite{GH,BCK}. It is pointed out that the spintessence model of
complex scalar may lead to formation of Q-balls, a drawback for dark
energy \cite{Kasuya}.

In this paper, we show that a real scalar field in semiclassical
 gravity may share some of useful properties of spintessence without
forming Q-balls. In semiclassical gravity approach a quantum real
 scalar field is equivalent to a complex scalar field with a
 constraint from the quantization rule \cite{BKKSY,Kim97,KJSS}.
The basic idea is as follows: the minimum uncertainty relation
satisfied by the field $\phi$ and the momentum $\pi_{\phi}$ [in
units of $c = \hbar =1$],
\begin{eqnarray}
\Delta \phi \Delta \pi_{\phi} \sim \frac{m_P^3}{2},
\end{eqnarray}
approximately leads to an effective potential
\begin{eqnarray}
V_{\rm eff} (\chi) \sim \frac{m_P^6}{8 a^6 \chi^2}
+ V (\chi),
\end{eqnarray}
where $\chi \sim \Delta \phi$. Here, the quantization rule or
uncertainty principle in semiclassical gravity plays the role of the
conserved charge of the spintessence model, the difference being
that the Planck constant (set to $\hbar = 1$) is a fundamental
constant in the former whereas the charge is an arbitrary parameter
in the latter.

Thus, the  minimum energy of the effective potential may be called
the quantum remnant of a dynamical scalar field. Note that the
physical argument for the quantum remnant is the uncertainty
principle. Any quantum field, contrary to the classical counterpart,
cannot settle down to the potential minimum  because fluctuations
originated from the uncertainty principle lead to a non-zero ground
state energy. Likewise, the same principle works in semiclassical
gravity for cosmology. The quantum real scalar field cannot rest at
the minimum of the classical potential, but it fluctuates around the
potential minimum.

We now study quantum effects of a single real scalar field in the
later evolution of the universe. In a spatially
 flat FRW universe with metric
\begin{eqnarray}
ds^2 = - dt^2 + a^2 (t) d{\bf x}^2,
\end{eqnarray}
the homogeneous real scalar field $\phi$ has the Hamiltonian
\begin{eqnarray}
\calh (t) = \frac{\pi_{\phi}^2}{2a^3} + a^3 V(\phi),
\end{eqnarray}
where $\pi_{\phi} = a^3 \dot{\phi}$. The equation of motion for
$\phi$ is given by
\begin{eqnarray}
\ddot{\phi} + 3 H \dot{\phi} + V^{\prime} = 0,
\label{homo_eom}
\end{eqnarray}
where $H = \dot{a}/a$ is the Hubble constant and a prime
denotes  derivative with respect to $\phi$.
 The homogeneous Friedmann equation is
\begin{eqnarray}
H^2 = \frac{8 \pi}{3 a^3 m_P} \calh.
\end{eqnarray}

However, in semiclassical gravity, the classical equations
are replaced by the Schr\"{o}dinger equation for $\phi$
\begin{eqnarray}
i \frac{\partial}{\partial t} \Psi(\phi, t) = \calh (t)
\Psi(\phi, t),
\label{schrodinger}
\end{eqnarray}
and by the semiclassical Friedmann equation
\begin{eqnarray}
H^2 = \frac{8 \pi}{3 a^3 m_P} \langle \calh \rangle_{\Psi}.
\end{eqnarray}
For a convex potential $V$, where the curvature is positive, a
generic Gaussian wave function may approximate the exact state near
the potential minimum \cite{BKKSY}:
\begin{eqnarray}
\Psi (\phi) = \frac{1}{(2 \pi \chi^2)^{1/4}} \exp
\Biggl[-\Biggr(\frac{1}{4\chi^2} - i \frac{\pi_{\chi}}{2 \chi}
\Biggr)\phi^2 \Biggr]. \label{gaussian}
\end{eqnarray}
Here, $\pi_{\chi} = a^3 \dot{\chi}$, and the auxiliary variables
satisfy the dispersions $\chi = \Delta \phi$ and $\pi_{\chi} =
\Delta \pi_{\phi}$, respectively. Now the Dirac action principle
leads to the effective action
\begin{eqnarray}
\calh_{\rm eff} = \frac{\pi_{\chi}^2}{2 a^3} + \frac{m_P^6}{8 a^6 \chi^2}
+ a^3 \langle V (\phi) \rangle. \label{eff ham}
\end{eqnarray}
The origin of the centrifugal potential, the second term, in Eq.
(\ref{eff ham}) is either the quantization rule or the uncertainty
principle. Hence, the semiclassical Friedmann equation reads
\begin{eqnarray}
H^2 = \frac{8 \pi}{3 a^3 m_P} \calh_{\rm eff},
\end{eqnarray}
and the field equation becomes
\begin{eqnarray}
\ddot{\chi} + 3 H \dot{\chi} - \frac{m_P^6}{4 a^6 \chi^3} +
\frac{\partial \langle V \rangle}{\partial \chi} = 0. \label{field}
\end{eqnarray}

In the oscillator representation, the Gaussian wave function is the
ground state of the annihilation and the creation operators that are
dimensionless,
\begin{eqnarray}
\hat{a} &=& \frac{\varphi^*}{m_P^3} \hat{\pi}_{\phi} - \frac{1}{m_P^3}
\pi_{\varphi}^{\ast}
\hat{\phi}, \label{anni_op}\\
\hat{a}^{\dagger} &=& \frac{\varphi}{m_P^3} \hat{\pi}_{\phi} - \frac{1}{m_P^3}
\pi_{\varphi} \hat{\phi},
\label{creat_op}
\end{eqnarray}
where $\pi_{\varphi} = a^3 \dot{\varphi}$ and a complex auxiliary
field $\varphi$ satisfies the equation of motion \bea \ddot{\varphi}
+ 3H \dot{\varphi} +V^{\prime} = 0. \eea The commutation relation
$[\hat{a},\hat{a}^{\dag}] = 1$, or the quantization rule yields the
condition
\begin{eqnarray}
\pi_{\varphi}^* \varphi - \varphi^* \pi_{\varphi} = i m_P^3. \label{con}
\end{eqnarray}
The auxiliary field $\chi$ is the amplitude of the complex scalar
field, $\varphi = \chi e^{- i \theta}$ and the quantization rule (\ref{con})
 becomes
\begin{eqnarray}
Q = 2 a^3 \chi^2 \dot{\theta} = m_P^3. \label{q-charge}
\end{eqnarray}
Note that Eq. (\ref{q-charge}) has the same form as the charge
conservation in the spintessence model \cite{GH,BCK}. The
centrifugal potential from the quantization, physically speaking,
the uncertainty principle, prevents the field $\chi$ from falling to
the classical minimum of $V$. Instead, it eventually
 reaches the limit cycle \cite{BKKSY}
\begin{eqnarray}
\frac{m_P^6}{4 a^6 \chi^3} = \frac{\partial \langle V \rangle}{\partial \chi}.
\label{limitcycle}
\end{eqnarray}

To investigate the later evolution of quantum remnants, we need to
check the equation of state. The energy density and the pressure for
the zero-mode of quantum scalar field are given by
\begin{eqnarray}
\rho &=& \frac{1}{a^3} \langle \calh \rangle =
\frac{1}{2}\dot{\chi}^2 + \frac{m_P^6}{8 a^6 \chi^2} + V(\chi), \\
p &=& \frac{1}{2}\dot{\chi}^2 + \frac{m_P^6}{8 a^6 \chi^2} - V(\chi).
\end{eqnarray}
Then the equation of state around the limit cycle is
\begin{eqnarray}
w = \frac{p}{\rho} \simeq \frac{\chi V^{\prime} - 2 V}{\chi
V^{\prime} + 2V},
\end{eqnarray}
where we have used Eq. (\ref{limitcycle}). The $w$ should be less
than $-1/3$ for dark energy. For a power-law potential, $V(\chi)
\sim \chi^q$, this constraint restricts $q<1$ \cite{BCK, Kasuya}.

For a general concave potential with negative curvature, we give a
simple argument based on the uncertainty principle. Roughly
speaking, the field operator and the momentum operator satisfying
the minimum uncertainty relation
\begin{eqnarray}
\chi \pi_{\chi} \sim \frac{m_P^3}{2},
\end{eqnarray}
approximately leads to an effective potential
\begin{eqnarray}
V_{\rm eff} (\chi) \sim \frac{m_P^6}{8 a^6 \chi^2}
+ V (\chi).
\label{effpot}
\end{eqnarray}
Therefore, the limit cycle is a consequence of the uncertainty principle.

\begin{figure}
\centering
\subfloat[Phase space]{\label{fig1a}
\includegraphics[width=0.35\textwidth]{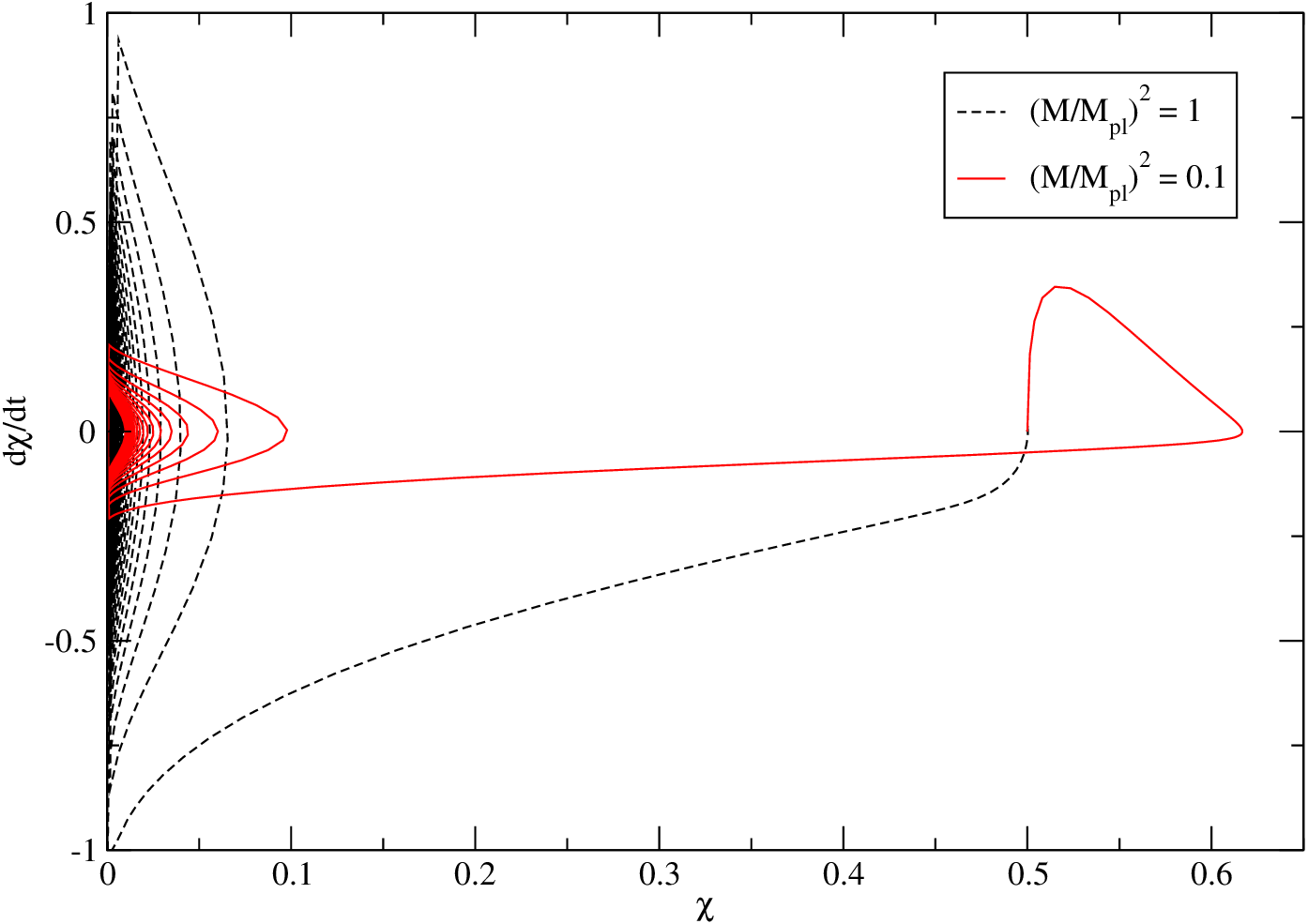}}
\subfloat[Evolution of equation of state parameter $w$]{\label{fig1b}
\includegraphics[width=0.35\textwidth]{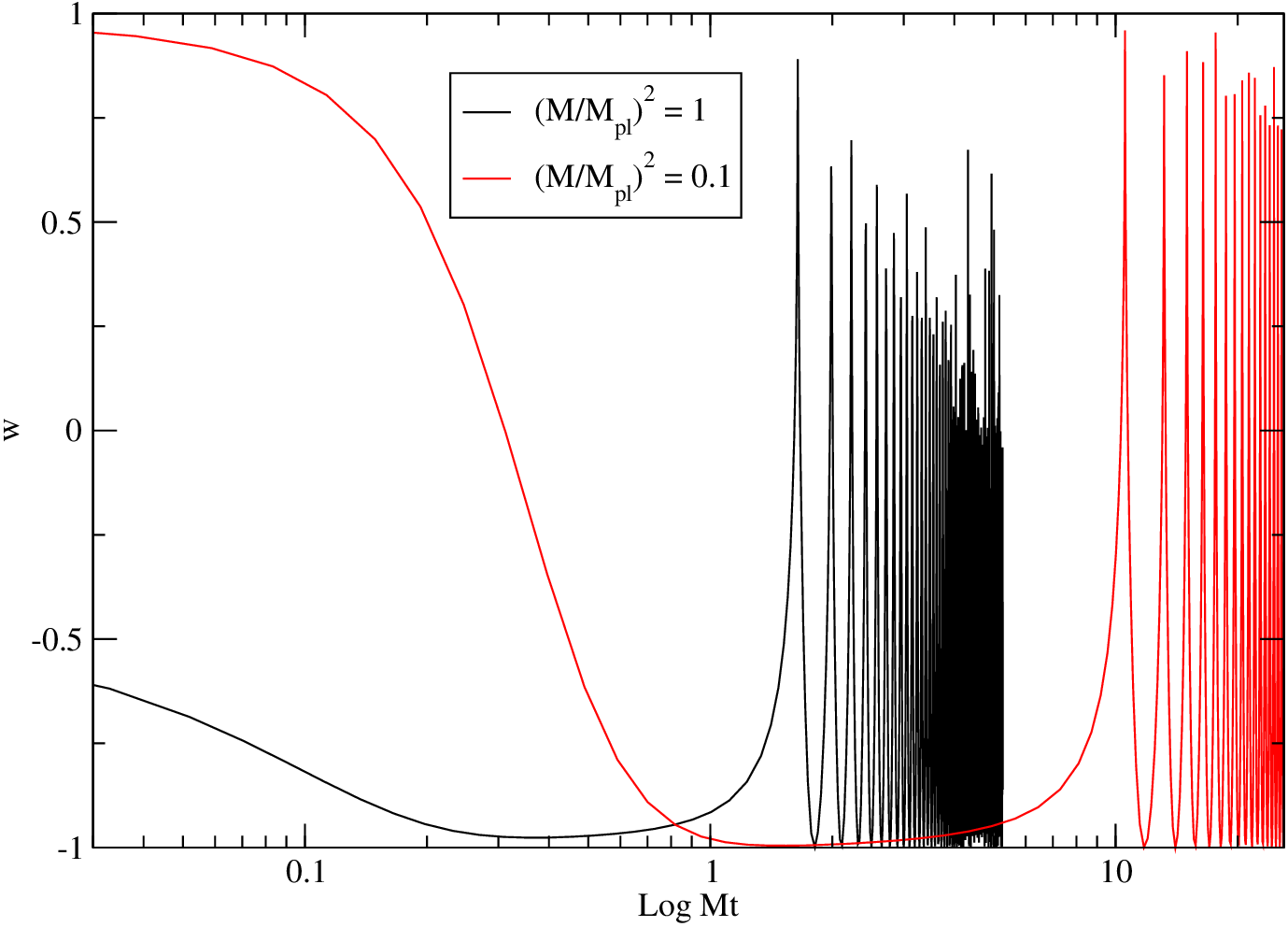}}
\caption{ (\ref{fig1a}) depicts the evolution of fields in phase space
 for $V = M^{4-q}\chi^q/q$
where $q=1/2$ and (\ref{fig1b}) plots the
evolution of equation of state parameter
$w$}
\label{fig1}
\end{figure}

\begin{figure}
\centering
\subfloat[Phase space]{\label{fig2a}
\includegraphics[width=0.35\textwidth]{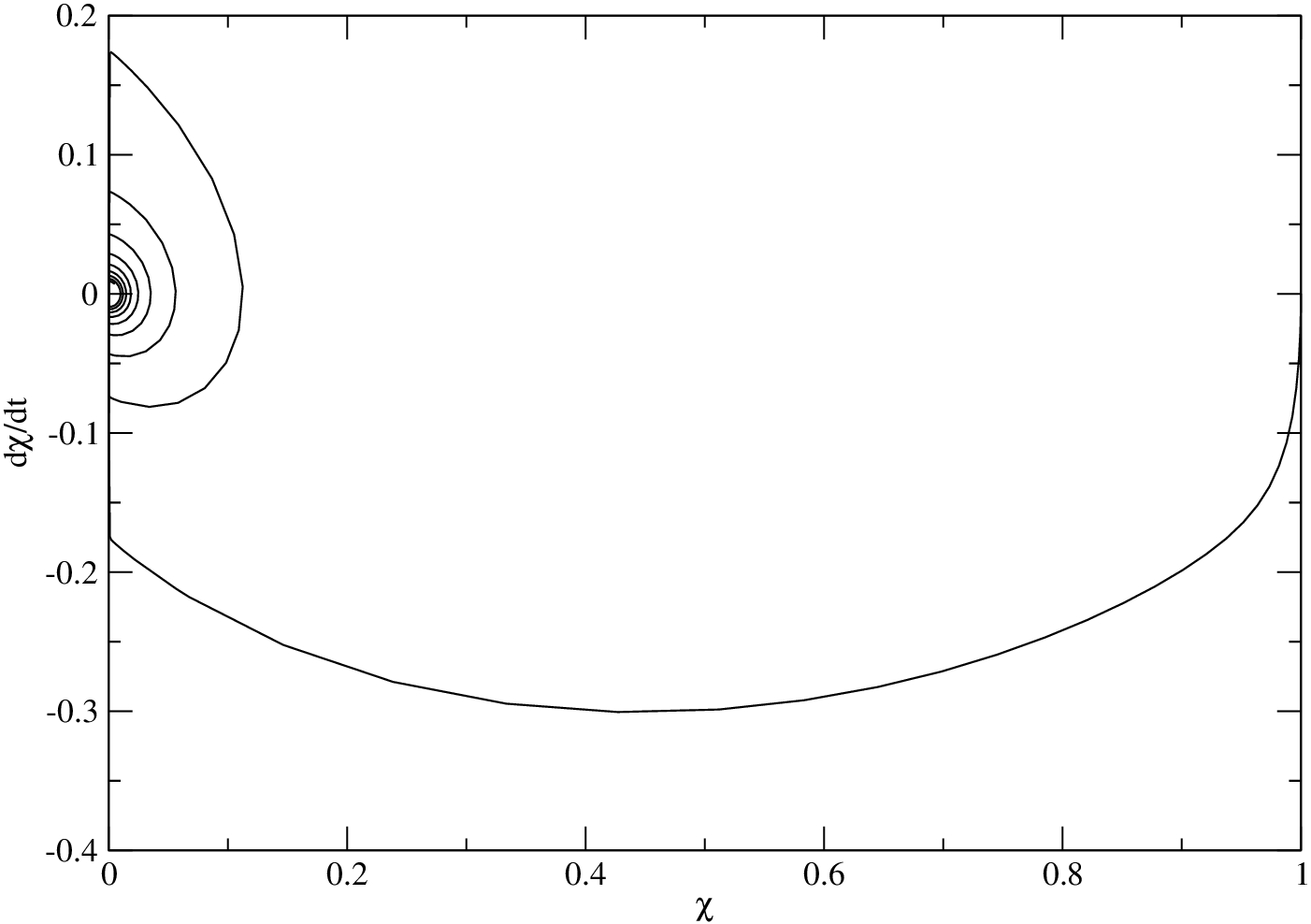}}
\subfloat[Evolution of equation of state parameter $w$]{\label{fig2b}
\includegraphics[width=0.35\textwidth]{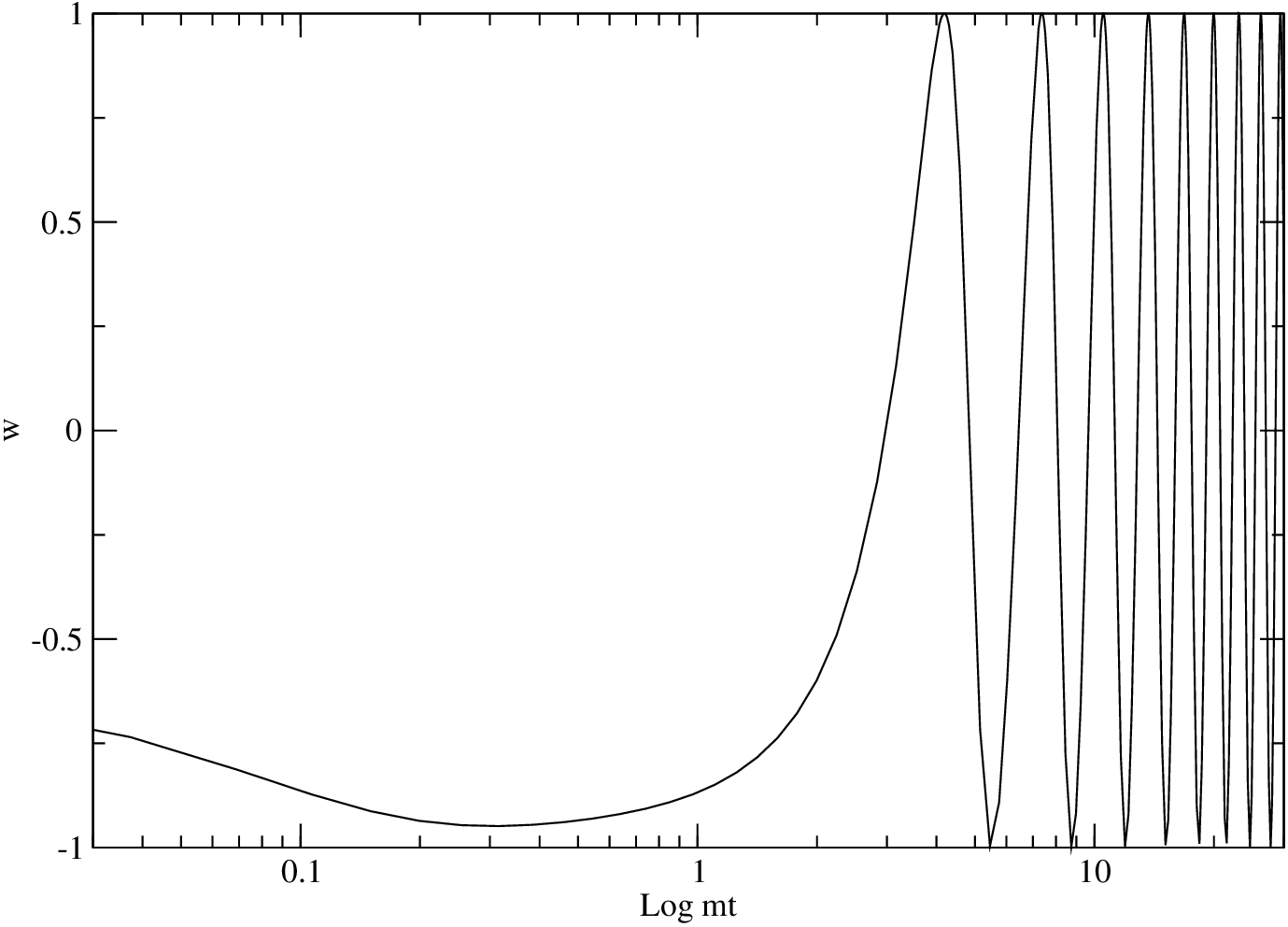}}
\caption{ (\ref{fig2a}) depicts the evolution of fields in phase space
 for $V = m^2 \chi^2/2$
with $m/m_{P} = 1$ and
(\ref{fig2b}) plots the evolution of equation of state parameter
$w$}
\label{fig2}
\end{figure}

In Figs. \ref{fig1} and \ref{fig2}, we plot the evolution of $\chi$
in  phase space
 and of  $w$ for $V(\chi) = 2M^{7/2}\chi^{1/2}$ and $m^2 \chi^2/2$,
respectively. It is numerically confirmed in Figs. \ref{fig1a} and
\ref{fig2a} that a later time, there is a barrier due to the
centrifugal potential term in Eq. (\ref{effpot}) which prevents
$\chi$ from falling to the potential minimum  and that $\chi$
approaches to the limit cycle at $\chi \simeq \chi_c$, where
$\chi_c$ is the value satisfying the condition (\ref{limitcycle}).
While $\langle w \rangle =0$ at limit cycle for $V(\chi) \sim
\chi^2$ in Fig. \ref{fig2b}, $\langle w \rangle < -\frac{1}{3}$ for
$V \sim \chi^{1/2}$ in Fig. \ref{fig1b} where we have performed time
average over equation of state parameter at limit cycle. As
discussed for a complex scalar field in Ref. \cite{BCK}, while the
power-law potential with $q=2$  can play the role of dark matter,
the power-law potential with $q<1$ can be considered as dark energy.

A few comments are in order. First, in discussing the Hamiltonian of
the quantum scalar field, we assumed its ground state to be
approximately described by
 the Gaussian wave function.
It is believed to be true in general for a convex potential,
especially $V\sim \phi^q$ with $q \geq 2$. However, for a (concave)
power-law potential with $q<1$ it may not hold because of a singular
behavior around $\phi =0$. In order to apply the approximate
Gaussian wave function for the power law potential with $q<1$, we
can consider the following form potential as in \cite{Damour:1997cb}
\begin{eqnarray}
V(\phi) = \frac{A}{q}\Biggl[ \Biggl(\frac{\phi^2}{\phi_c^2} +1
\Biggr)^{(1/2)q} -1 \Biggr], \label{osc_pot}
\end{eqnarray}
where $A$ is a dimensionful parameter. At  $\phi \gg \phi_c$, the
potential $V(\phi)$ can be approximated by the power-law potential
$V(\phi) \simeq \phi^{q}$, and if $\phi \ll \phi_c$, $V(\phi) \simeq
\phi^2$. Since $V'(\phi)$ is not singular any longer around $\phi=0$
even $q<1$, we may apply the Gaussian wave function approximately to
describe the ground state and take $\phi_c$ as the value that
satisfy the limit cycle condition (\ref{limitcycle}).

Next, we discuss the instability problem. The field $\phi(t, {\bf
x})$ is decomposed into the homogeneous part, $\phi_0(t)$, which
satisfies the equation of motion (\ref{homo_eom}), and fluctuations
$\delta \phi(t,{\bf x})$, which can be expressed in the Fourier
transform as
\begin{eqnarray}
\delta \phi(t,{\bf x}) = \int \frac{d^3k}{(2\pi)^{3/2}}[ \hat{a}
\varphi_k(t) e^{-i{\bf k}\cdot {\bf x}} + \hat{a}^{\dag}
\varphi_k^{\ast}(t) e^{i {\bf k}\cdot {\bf x}} ].
\end{eqnarray}
The expression $\hat{a}$ and $\hat{a}^{\dag}$ can be obtained from
Eqs. (\ref{anni_op}) and ( \ref{creat_op}). Then $\varphi_k$ satisfy
the equations of motion
\begin{eqnarray}
\ddot{\varphi}_k + 3H \dot{\varphi}_k + \Biggl(\frac{k^2}{a^2} +
V^{\dprime} \Biggr)\varphi_k = 0, \label{pert}
\end{eqnarray}
where we have neglected the metric perturbations. If $V^{\dprime}$
is assumed to include oscillating terms (for example, $V^{\dprime}
\approx V_0^{\dprime} + V_2^{\dprime}\cos 2\omega t$ and $\omega$ is
an oscillating frequency \cite{Johnson:2008se}), the above equation
reduces to the form of Mathieu equation \cite{Traschen:1990sw,
Kofman:1994rk, Kofman:1997yn}
\begin{eqnarray}
\frac{d^2}{dz^2}\varphi_k + (\mathcal{A}_k - 2\mathcal{B} 
\cos z)\varphi_k = 0,
\end{eqnarray}
where $z = \omega t, \mathcal{A}_k =
\frac{4}{\omega^2}(k^2+2V_0^{\dprime})$ and
 $\mathcal{B} = -\frac{2V_2^{\dprime}}{\omega^2}$
and we have ignored cosmic expansion. For a symmetric power-law
potential with $p\geq 2$, there exist stable solutions even in small
scale regions \cite{Taruya:1998cz}. For a power-law potential with
$q<1$, however, numerical calculations show that oscillating scalar
fields which have negative pressure
are unstable to the growth of inhomogeneities on small scales
\cite{Taruya:1998cz} and on large scales \cite{Johnson:2008se}. This
implies that instabilities are inevitable as far as the power-law
potential is concerned with $q<1$ which provides negative pressure.
In this sense such a potential may not be suitable for driving
cosmic acceleration.

However, the Gaussian wave function (\ref{gaussian}) may be not
appropriate to apply to the concave potential or potential
(\ref{osc_pot}) although we assume it would hold approximately in
the present work. We have to solve the Schr\"{o}dinger equation
(\ref{schrodinger}) for a concave potential or potential
(\ref{osc_pot}) and then improve the Gaussian wave function using a
variational method. And also in (\ref{pert}), we can take into
account higher order terms in $V^{\dprime}$
\begin{eqnarray}
\ \ddot{\varphi}_k + 3H \dot{\varphi}_k + \Biggl(\frac{k^2}{a^2} +
 \sum_{n=0}^{\infty} \frac{1}{2^n n!}\langle \varphi_k^2 \rangle
V^{(2n+2)}\Biggr)\varphi_k = 0,
\end{eqnarray}
where
\begin{eqnarray}
\langle \varphi_k^2 \rangle = \frac{1}{2\pi^2}\int dk k^2
|\varphi_k|^2,
\end{eqnarray}
and $V^{(n)} = d^n V/d\phi^n$. It is necessary to find the role of
higher order quantum corrections to the instability problem. We
shall leave these things to the future work.

In summary, we have studied the dynamics of a quantum real scalar
field that obeys the uncertainty principle and discussed the
possibility of its quantum remnant for dark matter and dark energy
at later evolution. It is shown that the uncertainty principle, by
preventing the quantum scalar field from rolling to the potential
minimum, can drive the cosmic acceleration in the recent era. In
fact, the uncertainty principle plays the similar role of the Hubble
friction in oscillating dark energy model or the centrifugal
potential in complex scalar fields (spintessence) model. If the
quantum remnants survived to the present era, those can play the
role of dark matter ($w=0$) or dark energy ($w<-1/3$) depending on
the shape of the potential. However, although no Q-balls are formed
in this model, quantum fluctuations are unstable to the growth of
the inhomogeneities for the power-law potential with $q<2$ which
provides negative pressure. As far as instability is concerned, the
model may not be suitable as a candidate for dark energy but is not
still excluded as a viable candidate for dark matter. So it would be
interesting to study whether the improved Gaussian wave function by
the variational method and thereby higher order quantum corrections
to the potential may resolve the instability problem or not.

\acknowledgments

S.~K. would like to appreciate the hospitality of Kunsan National
University, where this paper was completed, and Center for Quantum
Space Time (CQUeST) at Sogang University. This work was supported by
Korea Science Engineering Foundation (KOSEF) grant funded by the
Korea government (MOST) (No. F01-2007-000-10188-0).
\appendix

\end{document}